\documentclass{article}
\usepackage{INTERSPEECH2020}
\usepackage[utf8]{inputenc} %
\usepackage[T1]{fontenc}    %
\usepackage{url}            %
\usepackage{booktabs}       %
\usepackage{amsfonts}       %
\usepackage{nicefrac}       %
\usepackage{microtype}      %
\usepackage{lipsum,graphicx,xcolor}
\usepackage[normalem]{ulem}
\usepackage{multirow,array}
\usepackage{soul}
\usepackage{amsmath}
\DeclareMathOperator*{\argmin}{arg\,min}

\title{TTS Skins: Speaker Conversion via ASR}
\name{Adam Polyak$^{1,2}$\sthanks{ \hspace{0.1cm} The contribution of Adam Polyak is part of a Ph.D. thesis
research conducted at Tel Aviv University.}, Lior Wolf$^{1,2}$, Yaniv Taigman$^{1}$}
\address{
  $^1$Facebook AI Research \\
  $^2$The School of Computer Science, Tel Aviv University}
\email{adampolyak@fb.com}

\begin{document}
\maketitle

\begin{abstract}
We present a fully convolutional wav-to-wav network for converting between speakers' voices, without relying on text. Our network is based on an encoder-decoder architecture, where the encoder is pre-trained for the task of Automatic Speech Recognition, and a multi-speaker  waveform decoder is trained to reconstruct the original signal in an autoregressive manner. We train the network on narrated audiobooks, and demonstrate multi-voice TTS in those voices, by converting the voice of a TTS robot. %
\end{abstract}

\section{Introduction}

Text-To-Speech (TTS) and Automatic Speech Recognition (ASR) are key components in human-machine interfaces, deployed on popular virtual assistants. Recent advances in neural networks have dramatically increased the naturalness of TTS systems in synthesizing artificial human speech from text, thereby increasing their accessibility to users. 

These systems are trained end-to-end, from text to speech, on about 100 hours of a single speaker, usually recorded in a studio. Enriching this interface by providing additional speakers can either be done by collecting a similar dataset for each speaker, or through some adaptation process. Currently, most adaptation methods have shown limited success, either by requiring a similar data collection effort or by providing inferior results with respect to the identity of the speaker. 

We present a fully convolutional encoder-decoder network, that can be plugged into existing TTS systems and provide additional speaker identities (``skins''), requiring a considerably smaller quantity of target speaker recordings and without text.  Crucially, our encoder utilizes an ASR network that provides both high-level acoustic perception, as well as speaker-independent features. 

Our main contributions are: (i) We present the utility of pretrained ASR networks as suitable encoders for a voice conversion network. (ii) We present a conditioned WaveNet decoder based on these features. (iii) We propose a multispeaker TTS solution that has favorable properties, with regards to the required training data for the majority of the voices.

\section{Related Work}
\label{sec:related}

In addition to factors such as the expected quality and the computational complexity, a major factor in the applicability of a specific voice conversion technique is the type of supervision the method requires. Specifically, the quantity of the data that fitting a target speaker requires is of a great importance. Unlike many recent methods, our method does not require parallel training data between speakers and the reference source voice need not be available during training. The amount of data required varies between the phases of training. Since a high capacity WaveNet decoder is used, considerable (unlabeled) data is required for the training voices used for the initial training phase, in which the network and the look up table are being trained. Then, when fitting additional voices, much less data is required. 

The method of~\cite{liu2018wavenet} obtains data efficiency, by performing the conversion on a shared intermediate vocoder representation. This intermediate representation is estimated from linguistic features and decoded with a WaveNet. The WaveNet decoder is generic (from the vocoder features to waveform).  However, during fitting, both the part that maps the linguistic features to the vocoder features and the WaveNet are fitted to each new speaker. In contrast to our system, which uses an off-the-shelf ASR network, the linguistic features in~\cite{liu2018wavenet} are trained (on a proprietary dataset) together with the subsequent networks.

The topic of data efficiency was recently also studied in the supervised TTS case, where the training set contains both audio and text~\cite{chen2018sample}. Their conclusion, similarly to ours, is that it is beneficial to both optimize the embedding vector of the new speaker, as well as adopt the weights of the network itself.

Many of the literature voice conversion methods that require parallel data between the speakers, which is a limiting factor by itself, also require the parallel samples to be aligned in time. This requirement can be overcome in an iterative fashion, by first performing a nearest neighbor search of parallels and then refining the conversion module. Erro et al.~\cite{erro} have shown that repeating the matching process with the gradually improving conversion module leads to better and better results. A similar iterative process was used by Song et al.~\cite{song}, in which the transformation takes the form of adapting GMM coefficients. ASR features were used by Xie et al.~\cite{xie2016kl} in order to align multiple speakers in the latent space of the ASR system. Such an alignment relies on the ability of such features to be (mostly) speaker-agnostic. Once the alignment was formed, it was used in order to convert between voices, by concatenating matching fragments, i.e., in a unit-based approach.

In this work, we rely on a trained ASR network in order to obtain features. The ASR features are supposedly mostly orthogonal to the identity. An alternative approach would be to use the speaker identification features, in order to represent the source and target voices and perform the transformation between the two. I-vectors~\cite{ivector} are a GMM-based representation, which is often used in speaker identification or verification. Kinnunen et al.~\cite{7953215} have aligned the source and target GMMs, by comparing the i-vectors of the two speakers, without using transcription or parallel data. Unlike our method, the reference speaker is known at training time, and their method employs an MFCC vocoder, which limits the output's quality, in comparison to WaveNets. Speaker verification features were also used to embed speakers by Jia et al.~\cite{jia2018transfer}. In this case, the speaker embedding is based on a neural classifier.%

\begin{figure*}
\centering
\begin{minipage}[c]{0.7937\linewidth}
\includegraphics[trim=55pt 27pt 40pt 6pt, clip, width=0.99\linewidth, scale=0.5]{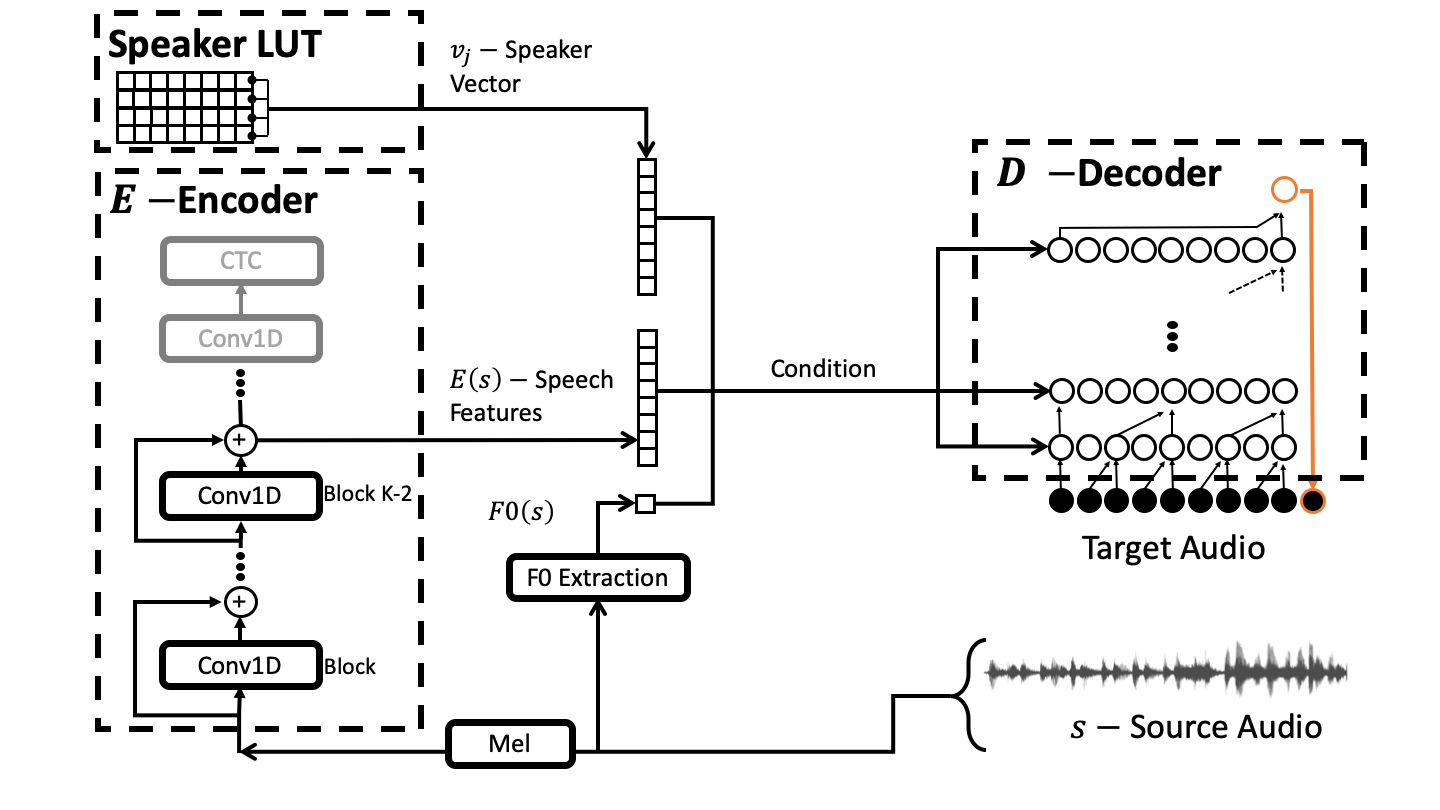}
\end{minipage}%
\hfill
\begin{minipage}[c]{0.19\linewidth}
\caption{\textbf{The architecture of our method.} The non-causal pre-trained encoder embeds the source audio signal, $s$, to speech features, $E(s)$. Then, the speaker's embedding, $v_i$, and the fundamental frequency values, F0(s), are concatenated to the speech features. The joint embedding is used to condition the autoregressive decoder, $D$, at multiple layers. During training, the decoder's input is the source audio, instead of the generated one. }
\label{fig:arch}
\end{minipage}
\end{figure*}

While in our method, the encoder is pre-trained for the purpose of ASR, other voice conversion methods employ a neural autoencoder. Specifically, many methods are based on variational auto encoders~\cite{vae}. Hsu et al.~\cite{hsu2016voice}, employs, similarly to us, one encoder and a parameterized decoder, where the parameters represent the identity. In a follow-up work~\cite{hsu2017voice}, a WGAN~\cite{wgan} loss term was added, in order to improve the naturalness of the resulting audio. In these variational autoencoder works, spectral frames are used, while our WaveNet decoder uses the waveform. 

A recent line of work~\cite{sun2016phonetic, saito2018non, zhao2018accent, Zhao2019ForeignAC} employs an ASR encoder to extract Phonetic Posterior Grams (PPGs), to perform voice conversion without parallel data. PPGs is a vector of probabilities for each phonetic class for a specific time-frame~\cite{hazen2009query}. Sun et al.~\cite{sun2016phonetic} predict STRAIGHT vocoder features from PPG features, assumed to be speaker independent, allowing them to perform many-to-one conversion. Saito et al.~\cite{saito2018non} present a method based on VAE for non-parallel voice conversion, conditioned on both speaker representations and phonetic content. Zhao et al.~\cite{zhao2018accent, Zhao2019ForeignAC} used PPG to perform accent conversion - first using conventional vocoders following with neural vocoders. Our work differs by (i) using  non-interpretable features rather than PPGs, and (ii) predicting wavform directly from those features rather than relying on spectral features (mel-spectrogram or vocoder features). Tian et al.~\cite{tian2019vocoder} presented a vocoder-free method for speaker conversion. Like us, they do not rely on intermediate spectral representation and directly feed phonetic context represented as PPGs to a WaveNet neural vocoder. Experimentally, we compare two different PPG representations to our method and show that our method leads to  more appealing audio samples.

An audio conversion method based on a waveform WaveNet autoencoder was presented in~\cite{mor2018autoencoderbased}. Trained mostly to convert between musical domains, voice conversion results are also presented. Unlike our method, the encoder is learned and multiple decoders are used, one per speaker. A single decoder that, similarly to ours, is conditioned on the identity of the target speaker was used in~\cite{NIPS2017_7210}. While, we employ a look up table that stores continuous speaker embeddings, their work employs a one-hot encoding. More importantly, the approach for obtaining a speaker-agnostic representation, is different. In~\cite{NIPS2017_7210}, the capacity of the latent space was reduced, by employing a discrete representation, while in~\cite{mor2018autoencoderbased}, a domain confusion loss is used. In our work, we employ neither approach, and instead rely on the speaker-agnostic nature of the ASR features.

\section{Method}
\label{sec:method}

The voice conversion method is based on training an autoencoder architecture, conditioned on an embedding of one of the $k$ speakers. In our method, the encoder is taken to be a subnetwork of a pre-trained ASR network, specifically the public implementation~\cite{jasper} of Wav2Letter~\cite{wav2letter}. Since an ASR network is used, by design, it is invariant to the speaker~\cite{adi2019reverse}, and the method does not require any disentanglement terms or other domain (speaker) confusion terms. This greatly adds to the stability of the method. During training, the speaker embeddings are learned and a softmax reconstruction loss is applied to the output of the shared WaveNet decoder.  

A diagram of the architecture is shown in Fig.~\ref{fig:arch}. It includes the pre-trained encoder $E$, the learned speaker embeddings and a single decoder, which is shared among all speakers, $D$.

The autoregressive decoder $D$ receives as input the sequence that was generated so far and is conditioned on: (i) the latent representation produced by the encoder $E$, and (ii) the desired speaker embedding. The first conditioning signal is obtained by upsampling this encoding in the temporal domain to obtain the original audio rate. Upsampling is done by repeating each conditioning frame to match the audio frequency. It is then applied, after the 128-dimensional speaker embedding is concatenated to it, to each of the WaveNet layers. 

\smallskip\noindent{\bf Training and the Losses Used.\quad} 
Since the encoder is taken off-the-shelf, we only train the decoder $D$. Let $D(\cdot,v)$ denote the conditioning of the decoder on a speaker that is embedded in a vector $v$. During training, we fit the vector $v_j$ of each speaker $j$ and store these in a Look Up Table (LUT). 
The following optimization problem is defined:
\begin{equation}
\min_{\theta_D,v_1,\dots, v_k}\sum_{j=1}^k\sum_{s^j} \mathcal{L}(D(E(s^j),v_j),s^j), 
\label{eq:loss}
\end{equation}
where $\theta_D$ are the parameters of the decoder, $s_j$ are the training samples of speaker $j$, $k$ is the number of speakers and $\mathcal{L}(o,y)$ is the cross entropy loss, applied to each element of the output $o$ and the corresponding element of the target $y$ separately. Note that the decoder $D$ is an autoregressive model that is conditioned on the output of $E$ in the previous frames. During training we use ``teacher forcing'', the autoregressive model is fed the target output $s^j$ from the previous time-step, instead of the generated output.

Once the network is trained, we can convert a new voice sample $s$ (of an existing or a new speaker) to an existing speaker $j$. This is done by applying the autoencoder pathway of speaker $j$, in order to obtain a conversion to this voice: $D(E(s),v_j)$.

Following Chen et al.~\cite{chen2018sample}, we present a variant of our method, which adds the fundamental frequency values $F0$ as additional conditioning to the network. {\color{black} $F0$ values were extracted at 5 ms steps using WORLD~\cite{morise2016world} vocoder. In the unvoiced regions, the values were set to a constant value across all speakers.} This enables the model to better preserve the prosody of the original voice sample $s$. Following this variant, the optimization problem is now defined as:
\begin{equation}
\min_{\theta_D,v_1,\dots, v_k}\sum_{j=1}^k\sum_{s^j} \mathcal{L}(D(E(s^j), v_j,F0(s^j)),s^j), 
\label{eq:loss2}
\end{equation}
where $F0(s^j)$ is the fundamental frequency series extracted from input voice sample $s$. Similarly, voice conversion for speaker $j$ is obtained via $D(E(s), F0(s),v_j)$.

\smallskip\noindent{\bf Fitting New Speakers.\quad}
\label{sec:fit}
It is desired to be able to add additional voices to the network post training, preferably with little training data and using a shorter training procedure. In order to achieve this, we perform fine-tuning of the pre-trained network, using the samples $s^{k+1}$ of the new speaker. In such a case, we fit both the embedding $v_{k+1}$ and fine tune the weights of the decoder $D$. Furthermore, instead of using randomly initialized embeddings for the new speakers, as was done in~\cite{jia2018transfer}, %
 the initial embedding of the new speaker $v^0_{k+1}$ is initialized with the best available embedding out of the training speakers. This is done by finding the speaker embedding which minimizes the loss in Eq.~\ref{eq:loss2} for samples $s^{k+1}$ of the new speaker:
\begin{equation}
v^0_{k+1} = \argmin_{v_1,\dots, v_k} \mathcal{L}(D(E(s^{k+1}), v_j,F0(s^{k+1})),s^{k+1}).  
\label{eq:embfit}
\end{equation}
After initialization, the loss of Eq.~\ref{eq:loss2} is used to refine the embedding $v_{k+1}$ and the decoder $D$, without reintroducing the training samples of the other speakers.

\smallskip\noindent{\bf Network Architecture.\quad}  
The encoder $E$ is a deep time delay neural network (TDNN), comprised of blocks of 1D-convolutional layers. It operates on log-mel filterbank extracted from a 16 kHz speech signal, using a sliding window of 20 ms with hops of 10 ms. Initially, the mel features sequence is downsampled by $\times 2$, using a single strided convolutional layer. The downsampeld input then passes a series of $10$ 1D-convolutional blocks, each with $5$ convolutional layers. The 1D-convolutional layers are composed of a sequence of a convolutional operation, batch normalization, clipped ReLU activation, and dropout. Blocks are composed of a series of the former layer, with a residual connection adding a projection of the block input to the block output. The model was trained on Librispeech~\cite{librispeech} together with data augmentation. We use the activations of layer \texttt{conv105} as input to the decoder, resulting a 768-dimensional conditioning signal at 50 Hz.
 
The WaveNet decoder $D$ has four blocks of ten dilated residual convolution layers~\cite{wavenet}. The layers in each block are with exponentially increasing dilation rate. We use 128 residual channels and 128 skip channels across all of the layers and a kernel-size of 2. At the top of the WaveNet, are two fully connected layers and a softmax activation, which provides probabilities for the quantized audio output (256 options) at the next timeframe. The autoregressive decoder generates audio at 24 kHz sampling frequency, resulting receptive field of 125ms (4,093 samples). We use 24 kHz sampling rate as it was shown to achieve higher subjective ratings~\cite{libritts}. In each WaveNet layer, the conditioning signal is passed to through a $1\times 1$ convolutional layer, that is unique for each conditioning location. 

We have developed efficient CUDA kernels that can decode autoregresively in real time (less than one second processing time for one second of output), thereby enabling efficient conversion. These kernels are loosely based on the nv-wavenet kernels~(\url{https://github.com/NVIDIA/nv-WaveNet}) provided by NVIDIA, and were optimized to utilize NVIDIA's Volta architecture. Additional modifications are based on the WaveNet variant of~\cite{nsynth} : (i) we add terms to ResNet skip connections that are directly computed from the WAV, and (ii) we add conditioning to the last fully connected layer.

\section{Experiments}

\begin{table*}[t!]
\centering
\caption{Test scores (Mean $\pm$ SD) for LibriTTS and VCTK. For MOS and identification, higher is better. For MCD -- lower.}
\label{tab:mos}
\begin{tabular}{@{}llc@{~}c@{~}cc@{~}c@{~}c@{}}
\toprule
\multirow{2}{*}{Data} & \multirow{2}{*}{Method} &  \multicolumn{3}{c}{LibriTTS} & \multicolumn{3}{c}{VCTK} \\
\cmidrule(l{2pt}r{2pt}){3-5} \cmidrule(l{2pt}r{2pt}){6-8}
 & &  MOS & MCD & Identification &  MOS & MCD & Identification\\
\midrule
Source voice & Ground truth & 4.45$\pm$0.68 & --- & 98.42 & 4.39$\pm$0.78 & --- & 99.5 \\
\midrule
\multirow{5}{*}{Seen during training}   & Full method &  \textbf{3.78$\pm$0.83} & --- & \textbf{96.12} & \textbf{4.08$\pm$0.75} & \textbf{8.76$\pm$1.72} & \textbf{98.97}  \\
                                        & w/o F0      &  3.61$\pm$0.83 & --- & 96.96 & 3.59$\pm$0.96 & 8.99$\pm$1.58 & 96.89 \\
                                        & Autoencoder baseline  &  2.89$\pm$0.88 & --- & 29.19 & 3.46$\pm$1.07 & 9.45$\pm$1.63 & 69.26 \\
                                        & PPG  &  2.82$\pm$0.91 & --- & 94.01 & 2.67$\pm$0.93 & 9.19$\pm$1.50 & 98.77 \\
                                        & PPG2  &  2.87$\pm$1.00 & --- & 95.77 & 3.03$\pm$1.06 & 9.18$\pm$1.52 & 96.24 \\
\midrule
\multirow{5}{*}{Unseen during training} & Full method &  \textbf{3.70$\pm$0.80} & --- & \textbf{97.10}          & \textbf{4.05$\pm$0.74} & \textbf{8.94$\pm$1.53} & \textbf{98.33} \\
                                        & w/o F0      &  3.67$\pm$0.82 & --- & 97.15          & 3.62$\pm$0.99 & 9.25$\pm$1.62 & 95.69 \\
                                        & Autoencoder baseline &  3.02$\pm$0.89 & --- & 32.55 & 3.83$\pm$0.91 & 9.65$\pm$1.51 & 66.20 \\
                                        & PPG  &  2.79$\pm$0.93 & --- & 94.05                 & 2.89$\pm$0.93 & 9.45$\pm$1.45 & 97.45 \\
                                        & PPG2  &  2.71$\pm$0.93 & --- & 95.43                & 3.19$\pm$1.04 & 9.79$\pm$1.86 & 97.25 \\
\midrule
\multirow{6}{*}{TTS robot} & Unconverted &  4.25$\pm$0.77 & 10.12$\pm$1.27 & ---              & 4.37$\pm$0.80 & 14.52$\pm$2.40 & --- \\
                           & Full method &  \textbf{3.67$\pm$0.81} & \textbf{8.13$\pm$0.95}  & \textbf{96.06}            & \textbf{4.17$\pm$0.88} & \textbf{12.68$\pm$2.17} & \textbf{99.25} \\
                           & w/o F0      &  3.47$\pm$0.76 & 8.43$\pm$0.97  & 96.66            & 3.75$\pm$1.07 & 13.06$\pm$2.26 & 96.36 \\
                           & Autoencoder baseline  &  3.02$\pm$0.84 & 9.38$\pm$1.09 & 60.26   & 3.85$\pm$1.05 & 13.81$\pm$2.29 & 75.56 \\
                           & PPG  &  2.91$\pm$0.94 & 8.52$\pm$0.93  & 96.63                   & 3.50$\pm$0.83 & 12.45$\pm$1.92 & 98.36 \\
                           & PPG2  &  2.85$\pm$0.87 & 8.76$\pm$1.06 & 95.08                   & 3.66$\pm$1.03 & 12.57$\pm$2.10 & 97.62 \\
\bottomrule
\end{tabular}
\vspace{-.3cm}
\end{table*}

\begin{table}[t]
\centering
\caption{MOS and similarity scores for the 2018 Voice Conversion Challenge (Mean $\pm$ SD; higher is better)}
\label{tab:mcdsingle}
\begin{small}
\begin{tabular}{@{}lcccc@{}}
\toprule
                    & \multicolumn{2}{c}{Hub} & \multicolumn{2}{c}{Spoke}  \\
\cmidrule(l{2pt}r{2pt}){2-3} \cmidrule(l{2pt}r{2pt}){4-5}
Method    			& MOS & Similarity & MOS & Similarity  \\
\midrule
GT source                   & 4.36$\pm$0.64 & 1.72$\pm$1.04 & 4.35$\pm$0.56 & 1.68$\pm$1.09 \\
GT target                   & 4.36$\pm$0.66 & 3.69$\pm$0.63 & 4.42$\pm$0.56 & 3.72$\pm$0.59 \\
N10~\cite{liu2018wavenet}   & 3.92$\pm$0.75 & 2.83$\pm$1.20 & 3.98$\pm$0.52 & 3.13$\pm$0.97\\
N17~\cite{wu2018nu}          & 3.27$\pm$0.95 & 2.77$\pm$1.17 & 3.40$\pm$0.88 & 3.05$\pm$0.96\\
\midrule
Ours                        & 3.84$\pm$0.85 & 2.87$\pm$1.14 & 4.00$\pm$0.55 & 3.14$\pm$0.97\\
\bottomrule
\end{tabular}
\end{small}
\vspace{-.4cm}
\end{table}

We present a series of experiments evaluating our method on: (i) many-to-many natural speech conversion on two public datasets, LibriTTS~\cite{libritts} and VCTK~\cite{vctk} from speakers seen during training, (ii) many-to-many natural speech conversion on the public datasets from speakers unseen during training, (iii) the customization of an existing off-the-shelf state of the art TTS engine~\cite{googlerobot} with identities from the public datasets and (iv) learning new identities with limited amount of data (less than five minutes) on the voice conversion challenge 2018~\cite{vcc2018}. Our samples can be found at~\url{https://tts-skins.github.io}, as well as in the supplementary.

LibriTTS is derived from the LibriSpeech corpus~\cite{librispeech} by removing noisy recordings, splitting sentences and 24kHz sample rate. We use speakers with at least five minutes of recordings from the 'train-clean-100' partition, as a result 199 speakers remain. In addition, we use the 'test-clean' partition, which includes 39 additional speakers, to evaluate our method on conversion of speakers unseen during training. The VCTK dataset is downsampled to 24kHz and eleven speakers are held-out for unseen speakers evaluation. Both datasets are split to 80\% train, 10\% validation and the rest is used for testing. On each dataset, we train a single ASR-features-conditioned WaveNet decoder. 

\smallskip\noindent{\bf Speaker conversion experiments.\quad}
We use both user-study and automatic based success metrics: (i) Mean Opinion Scores (MOS), on a scale between 1--5, which were computed using the crowdMOS~\cite{crowdmos} package. (ii) Mel cepstral distortion (MCD) between synthesized utterances and the reference utterances of the same speaker. Dynamic time warping is used to calculate the distortion for unaligned sequences. (iii) Speaker classification is employed to evaluate the capability of the method to synthesize distinguished voices. Similar to~\cite{deepvoice2,taigman2017voice,nachmani2018fitting}, automatic speaker identification results are obtained by training a multi-class CNN on the ground-truth training set of multiple speakers, and tested on the generated ones. The network operates on the WORLD vocoder features~\cite{morise2016world} extracted from the input and employs five batched-normalized convolutional layers of $3\times3$ filters with 128 ReLU activated channels. This is followed by max-pooling, average pooling over time, two fully-connected layers, and ending with a softmax corresponding to the number of speakers in the dataset.
Tab.~\ref{tab:mos} depicts the results of our method.

In experiments (i-iii), we compare with three baselines. The first is the WaveNet autoencoder voice conversion system of~\cite{mor2018autoencoderbased}. The second is a speaker conversion via PPG features marked as 'PPG'. PPG features were extracted using a $p$-norm DNN~\cite{zhang2014improving}, as done by~\cite{Zhao2019ForeignAC}. Finally, an additional PPG based speaker conversion system is trained by using a different set of PPG features, extracted by using pytorch-kaldi~\cite{ravanelli2019pytorch} framework. The underlying ASR model achieves an accuracy of PER=14.9\% on the TIMIT dataset~\cite{timit}. It is marked as 'PPG2'.

As can be seen, our full method improves the metrics over the baselines in all three settings. Furthermore, adding $F0$ improves the quality of our method output. Additionally, on the VCTK dataset, it also improves the identifiability of the generated samples. Finally, the quality of samples generated from voices seen in the training set is preferred over samples generated by converting samples from unseen speakers.

\smallskip\noindent{\bf The voice conversion challenge 2018.\quad}
VCC2018 evaluated 23 voice conversion systems on two tasks. First, the Hub task, in which  participants were asked to convert between speakers with parallel training corpora. Second, the Spoke task, for which participants were given training data for source and target speakers that contained different sets of utterances. Each task included four source speakers and four target speakers. Source speakers of the Hub task were all different from the source speakers in the Spoke task. Both tasks shared the same group of target speakers. The training data for each speaker (either source or target) is composed of 81 samples of recorded speech, the total amount of time varies between 4-5 minutes. 

In each task, participants are required to transform 35 utterances from each source speaker to each target speaker. As a result, each task requires the generation of $4\times4\times35=560$ samples. Samples are evaluated subjectively for their quality and similarity to a reference samples recorded by the target speakers. In our experiments, we used MOS for both quality and similarity evaluation. Quality of generated samples was rated on a scale of 1--5. Similarity was rated on a scale of 1--4, representing a binary similarity value (similar or not) and two values for certainty (sure/unsure) which included the amount of confidence. 
Following the challenge, the organizers published the submitted samples making it a suitable test bed for evaluating our method's ability to adapt to new speakers based on a limited amount of data, as described above in Sec.~\ref{sec:fit}.

Tab.~\ref{tab:mcdsingle} shows ours results versus the best performing method of the VCC 2018, which is known as N10~\cite{liu2018wavenet}. In addition, we included the results of N17~\cite{wu2018nu}, which matched the similarity scores of N10 on the Hub task. As can be seen, our system performs directly on par with the best performing system of the VCC 2018.

\section{Discussion}
This work enables customization of text-to-speech engines to an unlimited number of speakers. We show that combining a neural encoder obtained from an automatic speech recognition network with a neural audio decoder, achieves promising results. The alternative methods struggle to disentangle the speaker identity at the encoder level, while an ASR based encoder is trained to be speaker-agnostic. 

In addition to enabling multiple voice persona, the new technology can be readily used to create voice effects, e.g., convert ones voice to a whimsical voice. Being able to generate a large variety of voices is also beneficial to the study of identifying manipulated audio.

\clearpage
\bibliographystyle{IEEEtran}
\bibliography{references}

\end{document}